\documentclass[usenatbib]{mn2e_letter}
\usepackage[pdftex]{graphicx}
\usepackage{fixltx2e}
\usepackage{url}
\usepackage{amsmath,mathtools,revsymb,amssymb}
\usepackage{booktabs,multirow,calc,xspace,rotating,paralist}

\usepackage{etoolbox}
\makeatletter

\patchcmd{\NAT@citex}
  {\@citea\NAT@hyper@{\NAT@nmfmt{\NAT@nm}\NAT@date}}
  {\@citea\NAT@nmfmt{\NAT@nm}\NAT@hyper@{\NAT@date}}
  {}
  {}

\patchcmd{\NAT@citex}
  {\@citea\NAT@hyper@{%
     \NAT@nmfmt{\NAT@nm}%
     \hyper@natlinkbreak{\NAT@aysep\NAT@spacechar}{\@citeb\@extra@b@citeb}%
     \NAT@date}}
  {\@citea\NAT@nmfmt{\NAT@nm}%
   \NAT@aysep\NAT@spacechar%
   \NAT@hyper@{\NAT@date}}
  {}
  {}

\patchcmd{\NAT@citex}
  {\@citea\NAT@hyper@{%
     \NAT@nmfmt{\NAT@nm}%
     \hyper@natlinkbreak{\NAT@spacechar\NAT@@open\if*#1*\else#1\NAT@spacechar\fi}%
       {\@citeb\@extra@b@citeb}%
     \NAT@date}}
  {\@citea\NAT@nmfmt{\NAT@nm}%
   \NAT@spacechar\NAT@@open\if*#1*\else#1\NAT@spacechar\fi%
   \NAT@hyper@{\NAT@date}}
  {}
  {}

\makeatother

\usepackage[dvipsnames]{xcolor}
\definecolor{halfgray}{gray}{0.55}
\definecolor{webgreen}{rgb}{0,.5,0}
\definecolor{webbrown}{rgb}{.6,0,0}
\usepackage[pdftex,hyperfootnotes=false,pdfpagelabels]{hyperref}
\hypersetup{%
    colorlinks=true, pdfstartview=FitV,%
    breaklinks=true, pdfpagemode=UseNone, pageanchor=true, pdfpagemode=UseOutlines,%
    plainpages=false,%
    bookmarksnumbered, bookmarksopen=true, bookmarksopenlevel=1,%
    hypertexnames=true, pdfhighlight=/O,%
    urlcolor=webbrown, linkcolor=RoyalBlue, citecolor=webgreen,%
    pdftitle={Temperature Gradients in Clusters},%
    pdfauthor={\textcopyright\ authors},%
    pdfsubject={},%
    pdfkeywords={},%
    pdfcreator={pdfLaTeX},%
    pdfproducer={LaTeX with hyperref}}

\defcitealias{Voit2003}{V03}

\newcommand*{\mfigure}{figure\xspace}
\newcommand*{\mfig}{fig.\xspace}

\newcommand{\ie}{i.\,e.\xspace}

\newcommand{\eg}{e.\,g.\xspace}

\newcommand{\cf}{cf.\xspace}

\newcommand\icm{\textsc{icm}\xspace}

\newcommand\igm{\textsc{igm}\xspace}

\newcommand\agn{\textsc{agn}\xspace}
\newcommand\mhd{\textsc{mhd}\xspace}

\newcommand\mti{\textsc{mti}\xspace}
\newcommand\nfw{\textsc{nfw}\xspace}

\newcommand\sz{SZ\xspace}

\let\oldhat\hat

\renewcommand{\hat}[1]{\oldhat{\boldsymbol{#1}}}

\newcommand{\ts}[1]{\ensuremath{t_{\text{#1}}}}

\newcommand{\partiald}[2]{%
  \frac{\partial {#1}}{\partial {#2}}%
}

\newcommand{\kb}{\ensuremath{k_{\text{B}}}}

\newcommand{\kappae}{\ensuremath{\kappa_{\text{e}}}}
\newcommand{\kappaeff}{\ensuremath{\kappa_{\text{eff}}}}
\newcommand{\chie}{\ensuremath{\chi_{\text{e}}}}

\renewcommand{\mp}{\ensuremath{m_{\text{p}}}}

\newcommand{\fb}{\ensuremath{f_{\text{b}}}\xspace}
\newcommand{\rs}{\ensuremath{r_{\text{sh}}}\xspace}
\newcommand{\Rs}{\ensuremath{R_{\text{sh}}}\xspace}

\newcommand{\vi}{\ensuremath{v_{\text{i}}}\xspace}

\newcommand{\acknowledgments}{\begin{small}
    \section*{Acknowledgments}\end{small}}

\title[Temperature Profiles in the ICM]{%
  What Sets Temperature Gradients in Galaxy Clusters?  Implications
  for non-thermal pressure support and mass-observable scaling
  relations.}

\newcommand\altaffilmark[1]{\textsuperscript{#1}}
\newcommand\altaffiltext[1]{\textsuperscript{#1}}

\author[McCourt, Quataert, and Parrish]{
  \parbox[t]{0.9\textwidth}{
    \raggedright
    Michael McCourt,\altaffilmark{1}\thanks{E-mail:mkmcc@astro.berkeley.edu}
    Eliot Quataert,\altaffilmark{1} \&
    Ian J. Parrish\altaffilmark{1,2}}
  \vspace*{6pt} \\
  \altaffiltext{1}{Department of Astronomy and Theoretical Astrophysics
    Center, University of California
    Berkeley, Berkeley, CA 94720} \\
  \altaffiltext{2}{Present address: Canadian Institute for Theoretical
    Astrophysics, 60 St. George Street, University of Toronto,} \\
  {Toronto, ON M$5$S $3$H$8$, Canada} \\
}

\date{Submitted to MNRAS, December 2012}

\begin{document}
\maketitle
\label{firstpage}
\begin{abstract}
  We present a spherically symmetric model for the origin and
  evolution of the temperature profiles in the hot plasma filling
  galaxy groups and clusters.  We find that the gas in clusters is
  generically not isothermal, and that the temperature declines with
  radius at large distances from the cluster center (outside the core-
  and scale radii).  This temperature profile is determined by the
  accretion history of the halo, and is not quantitatively
  well-described by a polytropic model.  We explain quantitatively how
  the large-scale temperature gradient persists in spite of thermal
  conduction and convection.  These results are a consequence of the
  cosmological assembly of clusters and cannot be reproduced with
  non-cosmological simulations of isolated halos.  We show that the
  variation in halo assembly histories produces a $\sim10$\% scatter
  in temperature at fixed mass.  On top of this scatter, conduction
  decreases the temperature of the gas near the scale radius in
  massive clusters, which may bias hydrostatic mass estimates inferred
  from x-ray and \sz observations.  As an example application of our
  model profiles, we use mixing-length theory to estimate the
  turbulent pressure support created by the magnetothermal instability
  (\mti): in agreement with our earlier \mhd simulations, we find that
  the convection produced by the \mti can provide $\sim5$\%
  non-thermal pressure support near $r_{500}$.  The magnitude of this
  turbulent pressure support is likely to be non-monotonic in halo
  mass, peaking in $\sim 10^{14.5} \, M_\odot$ halos.
\end{abstract}


\section{Introduction}
\label{sec:introduction}
X-ray observations of the hot, diffuse gas in galaxy groups and
clusters suggest negative temperature gradients at large radii
\citep{Leccardi2008,George2009,Simionescu2011}.  This observation is
somewhat surprising because negative temperature gradients are
susceptible to a convective instability known as the magnetothermal
instability, or \mti \citep{Balbus2001}.  Without a clear source of
free energy to maintain the convection, it seems unusual that clusters
should so uniformly be found in unstable states.  This observation
becomes even more surprising in clusters more massive than
$\sim$10\textsuperscript{14.5}$M_{\odot}$, where the timescale for
heat to diffuse through the \icm can be much shorter than the Hubble
time.  Left to their own devices, both conduction and convection tend
to erase temperature gradients, and one might expect them to make the
gas isothermal.  In this paper, we study how the assembly of clusters
creates large-scale temperature gradients and maintains them in spite
of convection and thermal conduction.

Our subject is not purely academic.  Systematic trends in temperature
gradients with mass and redshift may influence the conversion of
observable quantities (such as the x-ray surface brightness) to
thermodynamic quantities (gas density, pressure, etc.).  Understanding
the origin of the temperature gradients in clusters would also enable
us to calculate how the turbulence produced by the \mti depends on
halo mass or redshift.  These trends create systematic variations in
non-thermal pressure support and might affect current efforts to use
the cluster mass function to constrain cosmology
\citep{Allen2008,Shaw2010,Allen2011}.  Thus, while the temperature
profiles in galaxy clusters present an interesting puzzle in their own
right, understanding the processes which control them may also find
useful application in cluster cosmology.

Several studies, including \citet{Dolag2004}, \citet{Burns2010}, and
\citet{Ruszkowski2011} have begun to address the effects of conduction
and the \mti in the \icm using cosmological simulations.  These
results are computationally expensive, however, and can be difficult
to interpret.  For example, the \mti is expected to be a sub-dominant,
but significant, source of turbulence in clusters
(cf. \citealt{Lau2009} and \citealt{Parrish2012}).  Using cosmological
simulations to study its trends with mass and redshift would require
very careful calibration of other sources of turbulence.  Thus, we
feel that a simplified treatment which affords an intuitive
understanding of the results remains useful.

The temperature and entropy profiles in the \icm are related by
hydrostatic equilibrium.  Accordingly, this paper closely follows
earlier work by \citet{Tozzi2001} and \citet{Voit2003}, who study the
entropy profiles in clusters.  However, the processes controlling
temperature gradients in clusters are slightly more subtle than those
which determine their entropy profiles.  Radial variations in entropy
tend to be much larger than those in temperature, so small differences
in the entropy profiles translate to much larger differences in
temperature profiles.  For the same reason, thermal conduction has a
more pronounced effect on the temperature profile than on entropy and
we must include it in our analysis.  Thus, despite significant
similarities to both \citet{Tozzi2001} and \citet{Voit2003}, our
models represent a generalization of these earlier studies and we use
them to explore different astrophysical applications.
\citet{Komatsu2001} also present analytic models for temperature
profiles in clusters.  As we describe below, however, our method
differs significantly from theirs because we do not assume a
polytropic model for the gas; instead we directly calculate its
thermodynamic state.

We describe our general methodology in the next section, followed by
two sets of models.  We begin in section~\ref{sec:adiabatic} with an
idealized but intuitive model.  This section illustrates the key
process that sets temperature gradients in clusters, but is too
idealized to be directly compared with real clusters.  We generalize
this model in section~\ref{sec:conduction} to more accurately reflect
the formation histories and gravitational potentials of clusters.  We
also study the influence of thermal conduction on our results.  We
close in \S~\ref{sec:discussion} with a brief summary of our
conclusions and a description of our future plans to apply them to
observations and to more realistic simulations.

\section{Method}
\label{sec:method}
Following \citet{Tozzi2001} and \citet{Voit2003} (hereafter
\citetalias{Voit2003}), we model the cosmological assembly of a galaxy
cluster and use the evolving properties of its accretion shock to
calculate the thermal state of the gas in its \icm.  Then, by assuming
the \icm is in hydrostatic equilibrium, we determine its temperature
and pressure profiles.  We focus our attention on how the accretion
history of a halo influences its temperature profile, and on how this
temperature profile is later modified by thermal conduction.  As
discussed in section~\ref{sec:introduction}, we build on previous work
that has focused on cluster entropy profiles.

Since the dynamical timescale in clusters is typically shorter than
the timescales for either accretion or thermal conduction, we
approximate the dynamics of the gas by assuming that it is in
hydrostatic equilibrium at all times.  We note that this assumption of
strict hydrostatic equilibrium within the virial shock is not
quantitatively justified: cosmological simulations of cluster
formation \citep[\eg][]{Rasia2006,Nagai2007,Lau2009,Vazza2011}
consistently find $\sim$10--20\% turbulent pressure support driven by
mergers near the virial radius.  Our assumption of hydrostatic
equilibrium represents a first approximation and provides a simple
model with no free parameters.

Our assumption of spherical symmetry is also an approximation: galaxy
cluster halos are tri-axial and grow partly by accreting smaller
sub-halos.  The extent to which these properties influence the
temperature profiles in clusters is not entirely clear; including them
in our analysis would require cosmological simulations, however, and
would likely obscure our results.  Instead, we model clusters as
spherically symmetric and we assume that they grow primarily via
smooth accretion.  Though these approximations are not entirely
realistic, they are partially justified in \citetalias{Voit2003}, who
compare models based on smooth accretion with more realistic ones
based on hierarchical structure formation: the differences are modest
for high-mass halos and mostly reflect changes in normalization due to
clumping factors.

Each shell of material accreted by the cluster contains both dark and
baryonic matter.  We do not solve for the evolution of the dark
matter; instead, we assume that the dark matter virializes rapidly and
we parametrize it with a time-dependent fit to the potential (we
neglect the gravity due to the baryons).  We consider both simplified,
isothermal potentials (\S~\ref{sec:adiabatic}) and more realistic fits
to cosmological simulations (\S~\ref{sec:conduction}).

The baryons enter the cluster through a shock with a density
$\rho_{\text{b,i}}$ and velocity $v_{\text{i}}$ determined by the
accretion rate and by the depth of the potential:
\begin{align*}
  \rho_{\text{b,i}} \, \vi &= \frac{\fb}{4\pi} \frac{1}{\rs^2}
  \frac{\partial M_{\text{sh}}}{\partial t} \\
  \vi^2 &= \frac{2 G M_{\text{sh}}}{\rs} (1-\xi).
\end{align*}
In the above, $r_{\text{sh}}$ is the radius of the accretion shock,
$M_{\text{sh}}$ is the total mass contained within it, and
$f_{\text{b}} \approx 0.17$ is the cosmic baryon fraction.  The
parameter $\xi = \rs/r_{\text{ta}}$ parametrizes the strength of the
virial shock, and $r_{\text{ta}}$ is the ``turnaround'' radius, at
which the kinetic energy of the shell vanishes
\citep{Gunn1972}.\footnote{Note that identifying $\xi$ with
  $\rs/r_{\text{ta}}$ assumes that the baryons do not separate from
  dark matter before reaching the virial shock.  If this were not the
  case, the baryons would feel a time-dependent potential due to the
  dark matter.  \citet{Tozzi2001} find that this may introduce a
  $\sim$10\% correction to the infall velocity.} We assume that the
\igm is cold enough that the gas enters the cluster through a strong
shock (with an upstream Mach number $M \gg 1$).  Thus, the post-shock
density and pressure are given by:
\begin{align*}
  P_{\text{sh}} &= \frac{3}{4} \rho_{\text{b,i}}
                \left(\frac{4}{3} \vi\right)^2 \\
  \rho_{\text{sh}} &= 4 \rho_{\text{b,i}},
\end{align*}
where the factor of $4/3$ in front of the velocity transforms the
infall velocity from the frame of the cluster to the frame of the
shock.  Thus, the post-shock entropy ($K \equiv \kb T n^{-2/3}$) is:
\begin{align}
  K_{\text{sh}} = \frac{1}{3} \left(\frac{\fb}{4\pi}\right)^{-2/3}
  \left[
    \frac{G^4 M_{\text{sh}}^2 (\mu m_{\text{p}})^5}%
    {\left(\frac{\partial \ln  M_{\text{sh}}}{\partial t}\right)^2}
    (1-\xi)^4
  \right]^{1/3}  .\label{eq:shock-entropy}
\end{align}
Equation~\ref{eq:shock-entropy}, above, is identical to equation~8 in
\citetalias{Voit2003}.  As emphasized in \citetalias{Voit2003}, apart
from an overall normalization $\propto (M/\fb)^{2/3}$, the entropy
profile depends only on the accretion history and on the shock
strength~$\xi$.

The jump conditions above represent the conservation of mass,
momentum, and energy and thus reflect the \textit{total} pressure
behind the shock.  In general, however, the electron and ion pressures
may differ.  The difference may be significant in clusters because the
shock preferentially heats ions and because the timescale for
electrons and ions to equilibrate is long \citep{Fox1997}.  We don't
distinguish between electron and ion temperatures in our calculation
because the temperature difference does not effect hydrostatic
equilibrium and thus should not influence our solutions.  Moreover,
the simulations by \citet{Rudd2009} show that this temperature
difference is modest within the virial radius of the cluster.

Our assumption of a strong virial shock may not be valid for the
innermost shells of material, which accreted when the \igm was hotter,
and when the gravitational potential of the halo was shallower, than
they are today.  Hence, our model will not accurately reproduce the
gas profiles near the centers of clusters \citep{Tozzi2001}.  Many
other processes, including cooling, heating, and the formation of the
central galaxy also affect the structure of the core, however
\citep{Voit2011}.  We instead focus on the gas at larger radii, near
the virial radius.  Both the large mass and the long cooling time of
this gas enable us to ignore galaxy formation processes at smaller
radii \citep[\eg][]{Voit2003b,Voit2011}.

After undergoing the virial shock, the gas entropy evolves due to
thermal conduction.  Thermal conduction is highly anisotropic in the
\icm because the electron mean free path is much longer than its
gyroradius.  Anisotropic conduction renders the \icm unstable to the
\mti \citep{Balbus2001}, however, which generates turbulence and may
isotropize the magnetic field \citep{McCourt2011,Parrish2012}.  We
therefore parametrize thermal conduction through an effective
isotropic conductivity.  We introduce the effective conductivity
$\kappaeff \equiv f_{\text{Sp}} \kappae$, where $f_{\text{Sp}} \sim
1/3$ is a suppression factor due to the magnetic
field.\footnote{Though this approximation is suitable for our
  purposes, we note in passing that the use of an isotropic
  conductivity significantly alters other processes in the \icm, such
  as convection \citep{Balbus2001,Quataert2008} and thermal
  instability \citep{Sharma2010,McCourt2012,Sharma2012}, and is thus
  not suitable for more detailed dynamical studies.  Interestingly, a
  suppression factor $f_{\text{Sp}} \sim 1/3$ turns out to be
  appropriate even if the magnetic field is tangled on very small
  scales \citep{Narayan2001}.} Thus, the evolution of the entropy is
determined by the following energy equation:
\begin{align}
  \frac{d}{d t}\ln K = \frac{2}{3 P} \nabla \cdot
  \left(
    \frac{\kappaeff}{\kb}
    \nabla T
  \right) .\label{eq:conduction}
\end{align}
In the simplifying case that $\kappaeff \to 0$, the entropy of each
shell is a constant in time and equal to $K_{\text{sh}}$, given by
equation~\ref{eq:shock-entropy}.  In addition to the conductivity
$\kappa$, we also use the thermal diffusion coefficient $\chie \equiv
\kappae / (n_{\text{e}} \kb)$, which has units of
cm\textsuperscript{2}/s.

As mentioned above, anisotropic thermal conduction in the \icm drives
a convective instability known as the \mti.  This convection carries a
heat flux which should technically be added to
equation~\ref{eq:conduction}.  However, the convective heat flux is
smaller than the conductive flux by a factor of
\begin{align*}
  \frac{Q_{\text{conv}}}{Q_{\text{cond}}} \sim
  \sqrt{\frac{m_{\text{e}}}{m_{\text{i}}}}
  \left(\frac{H}{\lambda_{\text{e}}}\right)
  \alpha^3
  \left(H \frac{d \ln T}{dr}\right)^{3/2} ,
\end{align*}
where $m_{\text{e}}$ and $m_{\text{i}}$ are the electron and ion
masses, $H$ is the pressure scale-height, $\lambda_{\text{e}}$ is the
electron mean-free-path, and $\alpha$ is the mixing-length parameter
of the convection.  This ratio is small ($\sim~10^{-2}$) in the \icm,
enabling us to ignore the convective heat flux in
equation~\ref{eq:conduction} \citep[cf.][]{Parrish2008}.

Our spherically symmetric model is most easily represented in
Lagrangian coordinates with the enclosed mass as the independent
variable.  Thus, the continuity equation and the equation for
hydrostatic equilibrium take the form:
\begin{align}
  \mu \mp \frac{d r}{d M_{\text{b}}} &= \frac{1}{4\pi r^2}
  \left(\frac{K}{P}\right)^{3/5} ,\label{eq:mass} \\
  \frac{d P}{d M_{\text{b}}} &= - \frac{g}{4\pi r^2} ,\label{eq:hse}
\end{align}
where $M_{\text{b}}$ is the \emph{baryonic} mass contained within the
radius $r$, and we adopt the entropy $K$ and pressure $P$ as our
primary thermodynamic variables.  In this initial study, we ignore
sources of non-thermal pressure support (such as cosmic rays, magnetic
fields, or turbulence).  However, we discuss in
section~\ref{sec:discussion} our plans to self-consistently include
turbulence generated by the \mti and to compare these results with
cosmological simulations.

Equations~\ref{eq:shock-entropy}--\ref{eq:hse} completely specify our
model except for the mass accretion history $M_{\text{sh}}(t)$ and the
gravitational field $g(r,t)$ of the halo.  We assume that these are
determined by cosmology and are unaffected by the baryonic formation
of the cluster.  We describe a simple, idealized model for the halo
formation in section~\ref{sec:adiabatic} and a more realistic model in
section~\ref{sec:conduction}.

\section{Simplified Adiabatic Models}
\label{sec:adiabatic}
Before studying the consequences of thermal conduction, it is useful
to understand the `baseline' temperature profile generated by the
halo's mass accretion history.  Thus, we begin with adiabatic models
which ignore thermal conduction.  We further isolate the influence of
the accretion history by assuming the gravitational potential is
isothermal and that the accretion rate of the cluster is independent
of time: $M_{\text{sh}}(t) = M_0 \times t/t_0$.  As we show later,
this prescription contains only a single free parameter and is perhaps
the simplest nontrivial model of the process we wish to study.  Though
the results in this section cannot be directly applied to clusters,
they highlight some of the key physics determining the temperature
gradients in clusters, and will assist in our interpretation of the
more detailed models in section~\ref{sec:conduction}.

The next step in our model is to determine the shock strength $\xi$.
The most logical choice would be to calculate $\xi(M_{\text{b}})$ so
that the cluster was in hydrostatic equilibrium at every epoch; in
fact, this is formally required to use equation~\ref{eq:shock-entropy}
for the post-shock entropy.  In this section, however, we make the
simplifying assumption that $\xi(M_{\text{b}})$ is a constant.  Thus,
the models presented in this section are not entirely self-consistent.
Our goal in this section is only to obtain an intuitive understanding
of how the accretion rate of a halo influences its temperature
gradient.  We present more detailed models, with more accurate
results, in section~\ref{sec:conduction}.

Since we have assumed a solution for $\xi(M_{\text{b}})$ and that the
evolution of the gas is adiabatic, the entropy profile
$K(M_{\text{b}})$ is uniquely determined at all times by
equation~\ref{eq:shock-entropy}.  Thus, it suffices to solve
hydrostatic equilibrium only at the present epoch -- this solution
cannot depend on the state of the cluster at earlier times.  Hence, we
need not track the evolution of the cluster, and the equations
determining the state of the gas reduce to ordinary differential
equations.

\subsection{Method}
\label{subsec:adiabatic-method}
Before solving the equations of our model, we recast them in a more
versatile dimensionless form.  Since we have a system of ordinary
differential equations with the shock as one boundary, we
de-dimensionalize the equations in this section using the properties
at the \emph{shock radius}.  (Note that this differs from the usual
convention of using the virial radius of the underlying dark-matter
potential.)  We introduce the constants $\Rs$ and $M_0$, which
represent the shock radius of the cluster at the present time $t_0$
and the total mass enclosed within it.  (In an isothermal potential,
$\Rs$ and $M_0$ each differ from the virial radius and mass by a
factor of $2\xi$.)  We also define a dynamical time $\ts{dyn} \equiv
(G M_0/\Rs^3)^{-1/2}$, along with the spatial coordinate $x \equiv r /
\Rs$ and the Lagrangian mass coordinate $y \equiv M_{\text{b}} / (\fb
M_0)$.

We introduce the dimensionless gas variables $P_1$, $\rho_1$, and
$K_1$ via:
\begin{subequations}
\begin{align}
  \rho_{\text{b}} &\equiv \frac{\fb}{4\pi} \frac{M_0}{\Rs^3}
  \times \rho_1 \\
  P &\equiv \frac{\fb}{4\pi} \frac{G M_0^2}{\Rs^4}
  \times P_1 \\
  K_1 &\equiv P_1/\rho_1^{5/3} .
\end{align}\label{eq:adiabatic-nondim}
\end{subequations}
Thus, the equations for mass conservation and hydrostatic equilibrium
become:
\begin{subequations}
\begin{align}
  \frac{d x}{d y} &= \frac{(1-\xi)^{4/5}}{3^{3/5}}
                     \left(\frac{\ts{dyn}}{t_0}\right)^{-2/5}
                     \frac{y^{4/5}}{x^2 P_1^{3/5}} \label{eq:isothermal-mass} \\
  \frac{d P_1}{d y} &= -\frac{1}{x^3}, \label{eq:isothermal-press}%
\end{align}\label{eq:isothermal-system}%
\end{subequations}
which we solve subject to the boundary conditions at the shock:
\begin{subequations}
\begin{align}
  x(y = 1) &= 1\label{eq:isothermal-bc1} \\
  P_1(y = 1) &= \frac{4\sqrt{2}}{3} (1-\xi)^{1/2}
                \left(\frac{\ts{dyn}}{t_0}\right)\label{eq:isothermal-bc2}
\end{align}
and at the center:
\begin{align}              
  x(y = 0) &= 0.\label{eq:isothermal-bc3}%
\end{align}\label{eq:isothermal-bcs}%
\end{subequations}
With these definitions, the solution is independent of the parameters
$\fb$, $\mu$, $\Rs$, and $M_0$.  Furthermore, the system is
over-determined with three boundary conditions and two equations.  The
shock radius $\xi$ is therefore an eigenvalue which must be chosen to
meet the inner boundary condition in equation~\ref{eq:isothermal-bc3}.

Thus, the \textit{only} free parameter in our system of equations is
the ratio $\ts{dyn}/t_0$, and we expect to find a one-dimensional
family of models.  Since the average density of any dark matter halo
$\bar{\rho} \sim 200 \rho_{\text{crit}}$ is independent of mass in
hierarchical structure formation, we do not expect the dynamical
timescale $\ts{dyn}$ to vary strongly among clusters at any
cosmological epoch.  The ratio $\ts{dyn}/t_0$ thus measures the age
(or, equivalently, the assembly rate) of the halo.

We note that the following solution to
equations~\ref{eq:isothermal-system} and~\ref{eq:isothermal-bcs}:
\begin{align}
  P_1 = \frac{1}{2 x^2}, \quad
  x   = y, \quad
  \xi = \frac{1}{4}\label{eq:isothermal-analytic-soln}
\end{align}
exists when the assembly rate satisfies:
\begin{align*}
  \frac{\ts{dyn}}{t_0} &= \sqrt{\frac{3}{32}}.
\end{align*}
We describe the physical significance of this special, isothermal
solution in section~\ref{subsec:adiabatic-results}.

For other values of the assembly rate $\ts{dyn}/t_0$, we solve
equations~\ref{eq:isothermal-system} and~\ref{eq:isothermal-bcs}
numerically using a predictor-corrector method, and we solve the
eigenvalue problem for $\xi$ with a bisection search.  We avoid the
singularity in equations~\ref{eq:isothermal-mass}
and~\ref{eq:isothermal-press} by solving them on a logarithmic grid.
Since we cannot apply the boundary condition
equation~\ref{eq:isothermal-bc3} in the logarithmic coordinates, we
obtain an approximate boundary condition at a finite radius by
expanding the equation for hydrostatic equilibrium near $x = 0$.
Assuming that the temperature remains finite, this equation becomes:
\begin{align*}
  T \to T_{\text{vir}} \left(\frac{d \ln \rho^{-1/2}}{d \ln r}\right)^{-1} .
\end{align*}
Combined with the fact that $K \propto T \rho^{-2/3} \propto
y^{4/3}$, this implies that $\rho \propto r^{-2}$, and thus that
\begin{align}
  T &\to T_{\text{vir}},\;\text{and} \label{eq:isothermal-central-temp} \\
  y &\to \left(\frac{T_{\text{vir}}}{T_{\text{sh}}}\right)^{1/2}
        \left(\frac{3 \rho_{\text{sh}}}{\bar{\rho}}\right)^{1/3} x \label{eq:isothermal-central-m}
\end{align}
as $x \to 0$.  We use equation~\ref{eq:isothermal-central-m} as a
boundary condition for the eigenvalue problem and
equation~\ref{eq:isothermal-central-temp} to check the accuracy of our
integration.  The analytic solution in
equation~\ref{eq:isothermal-analytic-soln} also permits a more
pedantic test of our method.

\begin{figure}
  \centering
  \includegraphics[width=3.33in]{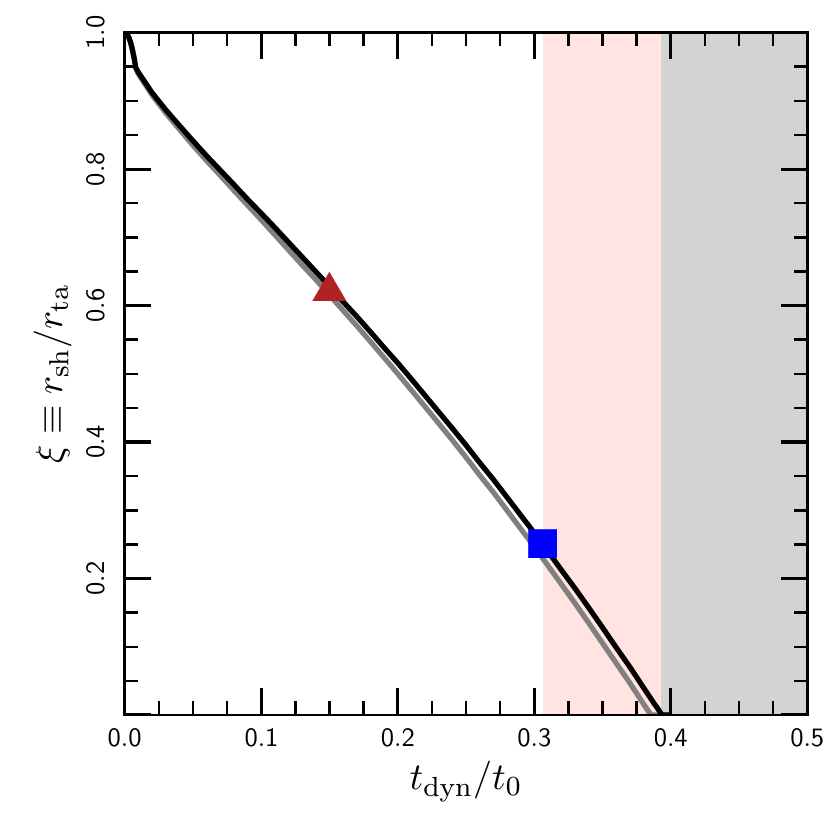}
  \caption{Shock radius $\xi$ as a function of the assembly parameter
    $\ts{dyn}/t_0$ for the simplified models described in
    section~\ref{sec:adiabatic}.  The timescale $t_0 = M/\dot{M}$ is
    the age of the halo.  Points further to the left on this plot
    correspond to clusters which form slowly and points further to the
    right correspond to clusters which form rapidly.  The red triangle
    represents the approximate value of $\ts{dyn}/t_0$ expected for
    cluster halos.  The blue square marks the isothermal solution in
    equation~\ref{eq:isothermal-analytic-soln}.  Clusters in the white
    region of the plot have negative temperature gradients, clusters
    in the pink region have positive temperature gradients, and
    clusters in the gray region cannot exist in our steady-state
    model.  Real clusters are expected to have moderate, negative
    temperature gradients based on this analysis (see
    \mfigure~\ref{fig:isothermal-temp-profiles}).  The black line
    illustrates the calculation with an isothermal potential, and the
    gray line shows results with a more realistic \nfw potential.}%
  \label{fig:isothermal-shock}
\end{figure}
%
\begin{figure}
  \centering
  \includegraphics[width=3.33in]{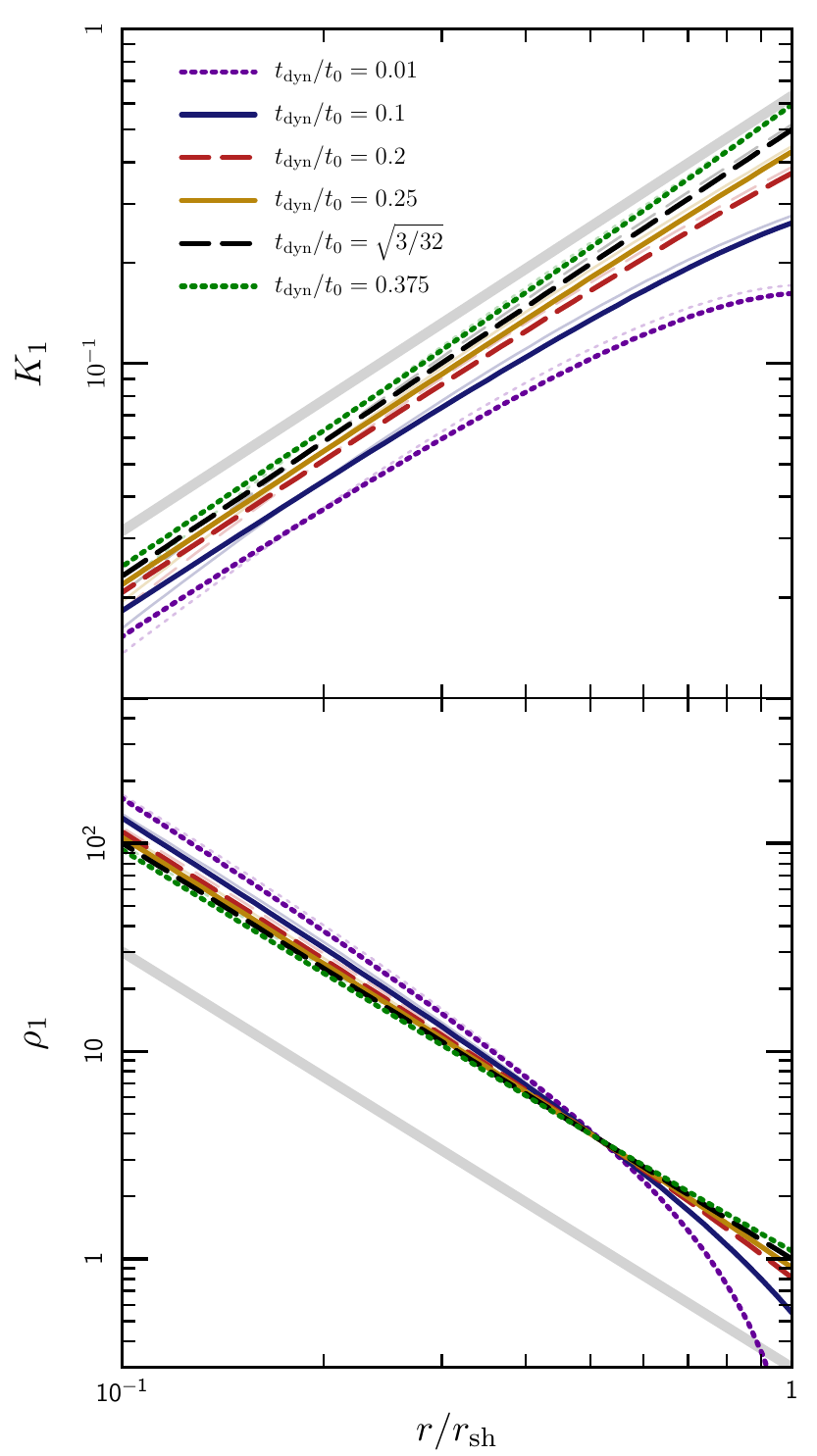}
  \caption{Profiles of density and entropy for representative models
    from \mfig~\ref{fig:isothermal-shock} with different values of the
    assembly parameter $\ts{dyn}/t_0$.  Thick gray lines show typical
    power-law slopes derived from x-ray observations
    \citep{Croston2008,Cavagnolo2009} (the normalization is
    arbitrary).  Massive clusters should mostly lie between the solid
    blue and long-dashed red lines.  Our solutions are approximately,
    but not exactly, power-laws.  The deviations from power-law
    behavior are dictated by the outer boundary condition on the
    pressure and determine the temperature profile
    (\mfig~\ref{fig:isothermal-temp-profiles}).  This boundary
    condition depends on the ram pressure behind the shock, and thus
    on the speed with which the cluster formed (parametrized by
    $\ts{dyn}/t_0$ in this model).  As in
    \mfig~\ref{fig:isothermal-shock}, thick curves show models with
    simplified, isothermal potentials and thin curves show more
    realistic models with \nfw potentials.}%
  \label{fig:isothermal-rho-ent-profiles}
\end{figure}

\subsection{Results}
\label{subsec:adiabatic-results}
Figure~\ref{fig:isothermal-shock} shows the solution for the
dimensionless shock radius $\xi(\ts{dyn}/t_0)$ from our numerical
calculations.  In the limit of very slow accretion (\ie as
$\ts{dyn}/t_0 \to 0$), the shock is comparatively weak and the
dimensionless shock radius $\xi \to 1$.  As the assembly rate
$\ts{dyn}/t_0$ increases, the shock radius $\xi$ decreases
monotonically.  Thus, the virial shock moves inwards as the accretion
rate increases.  This result seems intuitively reasonable, as a higher
accretion rate implies a higher ram pressure behind the shock.  A
stronger shock (or smaller $\xi$) is thus required to hold back the
infalling material and to keep the \icm in hydrostatic equilibrium.

Interestingly, \mfigure~\ref{fig:isothermal-shock} indicates a maximum
assembly rate around $\ts{dyn}/t_0 \approx 0.39$ at which $\xi \to 0$.
Beyond this point, thermal pressure alone cannot hold back the
accretion shock and hydrostatic equilibrium becomes impossible.  This
represents an extremely rapid accretion rate, however, with the halo
forming over only $\sim{}2.5$ dynamical times; our quasi-static model
for the \icm breaks down in this limit.  A fully dynamical calculation
(\eg a simulation) with such a high accretion rate would likely
produce a model \icm with significant time dependence and turbulent
pressure support, but with a finite shock radius.

Figure~\ref{fig:isothermal-shock} shows the isothermal solution
(eq.~\ref{eq:isothermal-analytic-soln}) as a blue square and a point
representative of a massive cluster [with $\ts{dyn} = 0.1 H_0^{-1}$
and $t_0 \sim (2/3) H_0^{-1}$] as a red triangle.  Our model predicts
that $\xi \sim 0.6$ for this fiducial cluster, similar to what has
been expected in the past \citep[\eg][]{Rees1977}.  Note that the
isothermal solution requires a much faster assembly than is typical
for galaxy clusters; this simple model thus suggests that clusters
should not be isothermal.

\begin{figure}
 \centering
 \includegraphics[width=3.33in]{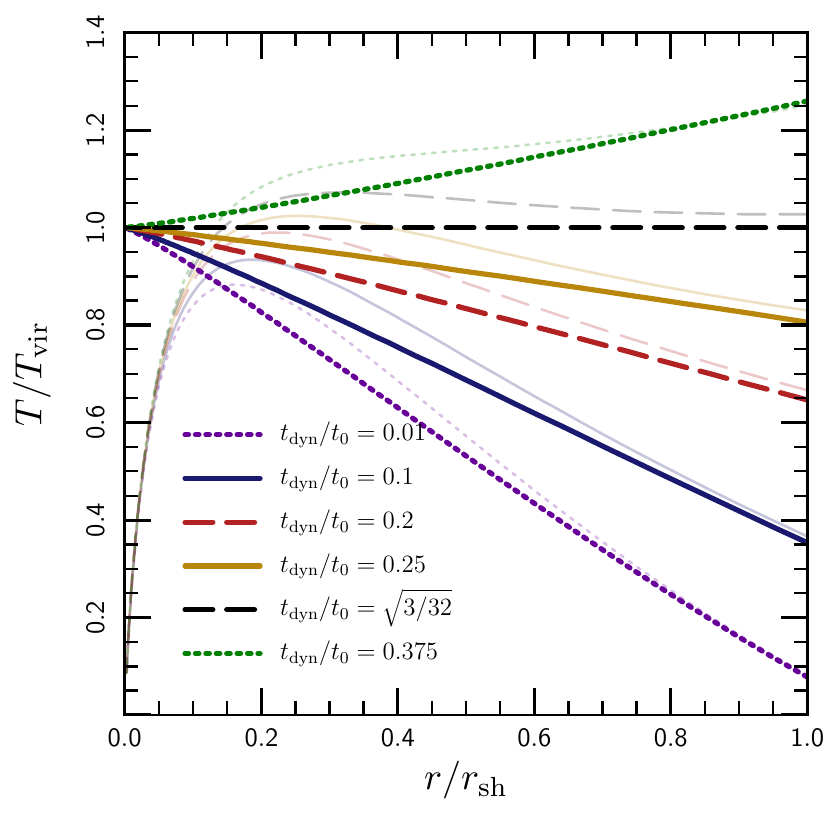}
 \caption{Temperature profiles for representative models from
   \mfig~\ref{fig:isothermal-shock} with different values of the
   assembly parameter $\ts{dyn}/t_0$.  The black line corresponds to
   the isothermal solution (eq.~\ref{eq:isothermal-analytic-soln});
   massive clusters should lie between the blue and red lines.  Our
   model implies temperature profiles which decrease by a factor of
   $\sim\,2$ from the center to the virial radius, in line with x-ray
   observations.  Thin, light lines show the results of calculations
   with \nfw potentials, rather than isothermal ones (see
   \S~\ref{sec:conduction} for details).  The temperature gradients
   are similar outside the scale radius of the halo.}%
 \label{fig:isothermal-temp-profiles}
\end{figure}
Figure~\ref{fig:isothermal-shock} directly illustrates the effect of
the accretion rate on the location of the virial shock.  The gas
properties in the \icm must match onto jump conditions at the shock;
thus, by moving the shock radius, the accretion rate also influences
the thermodynamic structure of the \icm.
Figure~\ref{fig:isothermal-rho-ent-profiles} demonstrates this by
showing profiles of the gas density and entropy for models with
different assembly rates $\ts{dyn}/t_0$.  The profiles are nearly
isothermal, with $\rho \sim r^{-2}$ and $K \sim r^{4/3}$, and are
broadly consistent with determinations from x-ray data
\citep{Croston2008,Cavagnolo2009}.

The small deviations from power-laws in the density and entropy
profiles lead to significant temperature gradients, however.  We show
this explicitly in \mfigure~\ref{fig:isothermal-temp-profiles}, where
we plot the temperature profiles for the models from
\mfigure~\ref{fig:isothermal-rho-ent-profiles}.  The temperature
profiles are nearly linear, and gradients of either sign are possible,
depending on the assembly rate $\ts{dyn}/t_0$.  This result is not
surprising, since the central temperature must equal the virial
temperature of the halo (eq.~\ref{eq:isothermal-central-temp}) and the
temperature at the shock is dictated by the jump conditions.  A
roughly linear interpolation between these boundary conditions seems
reasonable given the simplicity of the model.  The isothermal solution
with $\ts{dyn}/t_0 = \sqrt{3/32}$ divides models with negative and
positive temperature gradients.  In reality, most clusters satisfy
$\ts{dyn}/t_0 \sim 0.1-0.2$; in this case our model predicts
temperature profiles which decrease by a factor of $\sim 2$ from the
center to the shock radius.  This result is in line with recent
observations of the gas temperature near the virial radius
\citep{George2009,Simionescu2011}.

Thus, while we solve the system of equations~\ref{eq:isothermal-mass}
and~\ref{eq:isothermal-press} and boundary
conditions~\ref{eq:isothermal-bc1}--\ref{eq:isothermal-bc3} for the
shock radius and for the structure of the \icm, we find that the
boundary conditions essentially dictate the temperature profile.  The
temperature must reach the virial temperature of the potential at the
center of the halo, and it must match onto the jump conditions at the
shock.  The solution to hydrostatic equilibrium then implies a nearly
linear interpolation between these two boundary conditions.

As presented here, the temperature profiles in
\mfigure~\ref{fig:isothermal-temp-profiles} may seem specific to our
assumption of an isothermal potential.  To demonstrate that this is
not the case, we have included calculations with \nfw potentials in
\mfigure~\ref{fig:isothermal-temp-profiles} (shown as thin, light
lines).  Although the different potential has a dramatic effect on the
temperature profile within the scale radius, the trend between the
assembly rate and the overall temperature gradient at large radii is
similar.  We include these lines only for illustration, but present
much more detailed models with \nfw potentials in the following
section~\ref{sec:conduction}.

The simple model presented in this section suggests that the assembly
rate of the halo (or, equivalently, the ram pressure behind the
accretion shock) dictates the large-scale temperature gradient in the
\icm.  The assembly rates of massive clusters are such that they
should have moderate, negative temperature gradients outside the scale
radius.  This is one of our primary findings.  In what follows, we
show that this result holds true even when we relax the simplifying
assumptions in this section.  We also study how thermal conduction
modifies this `baseline' temperature profile.

\section{Conduction and Realistic Assembly Histories}
\label{sec:conduction}
Our models with isothermal potentials and linear accretion histories
are especially transparent.  The models are very idealized, however,
and it is not clear how accurately they carry over to real clusters.
In this section, we generalize our results to include more realistic
potentials and accretion histories; we also include thermal conduction
in our calculation and study its influence on the temperature profiles
in clusters.

\subsection{Method}
\label{subsec:conduction-method}
\subsubsection{Coordinates and Assumptions}
\label{subsubsec:conduction-coords}
Introducing realistic potentials and accretion histories into our
model necessitates a few changes to our method.  Though the unit
system introduced in section~\ref{sec:adiabatic} is ideal for our
model equations, the virial mass of the halo (as opposed to the mass
enclosed by the virial shock) is an eigenvalue of the problem and
cannot be specified ahead of time.  Since we want to study the
variation in temperature profiles at fixed \textit{virial} mass, we
must alter the equations and boundary conditions slightly.

We adopt a unit system based on the virial mass $M_{\text{vir},\,0}$
and radius $r_{\text{vir},\,0}$ of the halo at redshift zero.
Following our approach in the last section, we introduce the
dimensionless variables $P_1$, $\rho_1$, and $K_1$ via:
\begin{subequations}
\begin{align}
  \rho_{\text{b}} &\equiv \frac{\fb}{4\pi}
  \frac{M_{\text{vir},\,0}}{r_{\text{vir},\,0}^3}
  \times \rho_1 \\
  P &\equiv \frac{\fb}{4\pi}
  \frac{G M_{\text{vir},\,0}^2}{r_{\text{vir},\,0}^4}
  \times P_1 \\
  K_1 &\equiv \frac{P_1}{\rho_1^{5/3}} .
\end{align}\label{eq:conduction-nondim}
\end{subequations}
We also define the spatial coordinate $x \equiv r / r_{\text{vir}}$,
the Lagrangian coordinate $y \equiv M_{\text{b}} / (\fb
M_{\text{vir}})$, and the dynamical time $\ts{dyn} \equiv (G
M_{\text{vir}} / r_{\text{vir}}^3)^{-1/2}$.  Note that our definitions
of $x$, $y$, and $\ts{dyn}$ use instantaneous values of
$r_{\text{vir}}$ and $M_{\text{vir}}$, while our definitions of
$\rho_1$, $P_1$, and $K_1$ are normalized to $r_{\text{vir},\,\,0}$
and $M_{\text{vir},\,\,0}$.  Thermodynamic quantities in our
calculation (\eg $K_1$) are thus directly comparable at different
redshifts, while coordinates (\eg $x$) are not.

We define $m \equiv M_{\text{vir}}(t)/M_{\text{vir},\,0}$, which
tracks the formation of the halo and functions as a time coordinate.
We take the virial radius to be $r_{200}$, the radius within which the
mean density of the halo is $200$ times the critical density of the
universe.  Thus, $\ts{dyn} = (10 H)^{-1}$, where $H$ is the Hubble
parameter.  

Since the virial radius does not directly enter into our model
(\S~\ref{sec:adiabatic}), our choice defining the virial radius is
essentially arbitrary.  We chose $r_{200}$ because it is
straightforward to compute and facilitates comparison with much of the
existing literature.  We continue to use the notation
``$r_{\text{vir}}$'' over the notation ``$r_{\text{200}}$'' in order
to de-emphasize this arbitrary definition, however.

\subsubsection{Dark Matter}
\label{subsubsec:conduction-dm}
\begin{figure*}
  \begin{minipage}[t]{3.33in}
  \centering
  \includegraphics[width=3.33in]{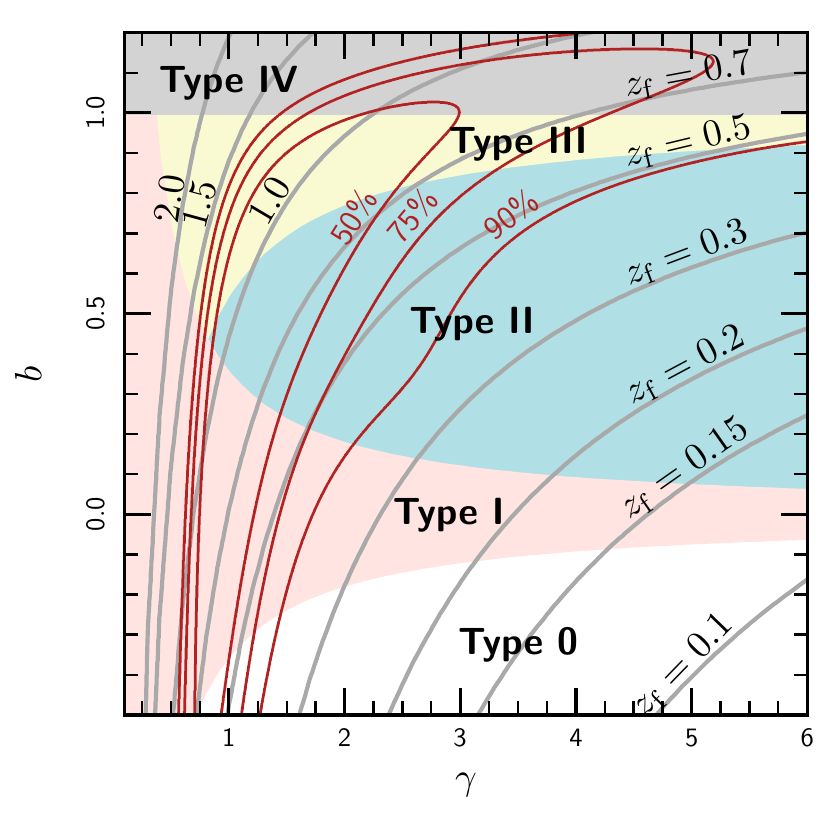}
  \caption{Parameter space for dark matter accretion histories based
    on the Millennium Simulation.  The exponents $\gamma$ and $b$ are
    defined in equation~\ref{eq:mdot-james}.  Colors demarcate the
    different accretion ``types'' from \citet{McBride2009} and gray
    lines show contours of the formation redshift $z_{\text{f}}$ such
    that $M_{\text{vir}}(z_{\text{f}}) = 0.5\,M_{\text{vir}}(z=0)$.  At
    redshift zero, Type~III halos are accreting very slowly and
    Type~$0$ halos are accreting rapidly.  The red curves mark
    contours of the probability density for $10^{15} M_{\odot}$
    clusters to have a given accretion history (taken from the
    Appendix of \citealt{McBride2009}).  The contours for $10^{14}
    M_{\odot}$ halos are very similar to those shown here.}%
  \label{fig:zf-contours}
  \end{minipage}
  \hspace*{\fill}
  \begin{minipage}[t]{3.33in}
  \includegraphics[width=3.33in]{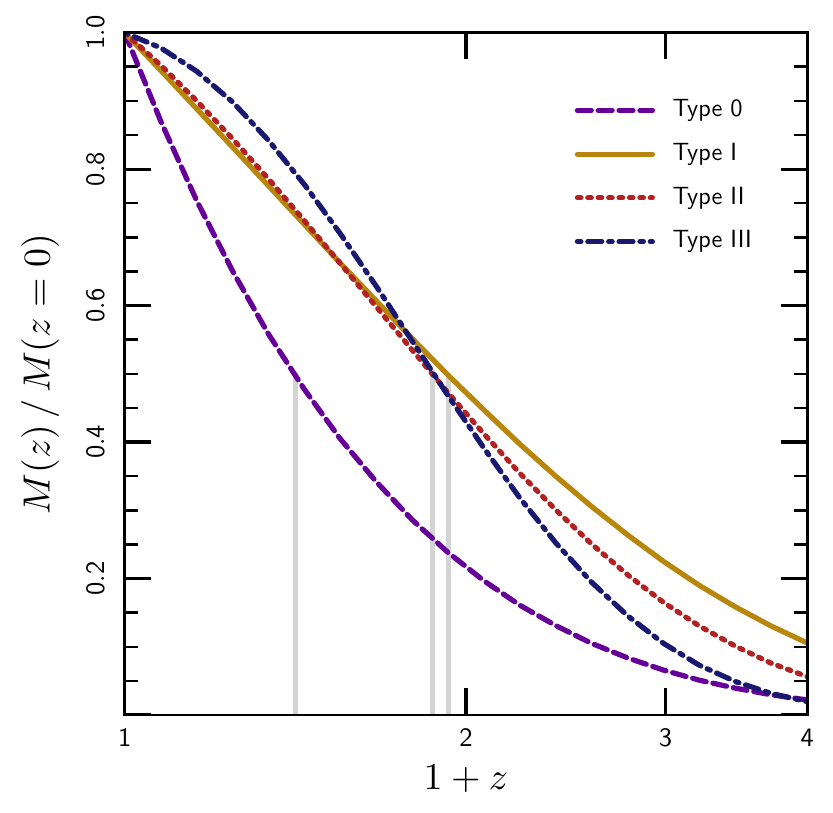}
  \caption{Example accretion histories from
    equation~\ref{eq:mdot-james} illustrating each of the four
    ``types.''  The examples correspond to $\{\gamma,b\} = $ $\{0.75,
    -1.5\}$ (Type~$0$), $\{0.75, 0.0\}$ (Type~I), $\{1.25, 0.5\}$
    (Type~II), and $\{2.25, 0.9\}$ (Type~III).  These examples are
    also used in figure~\ref{fig:temp-by-type}.  Note that Type~$0$
    halos have larger accretion rates, and that Type~III halos have
    lower accretion rates, than intermediate halo types at low
    redshift.  Vertical, gray lines indicate the ``formation
    redshift'' for which $M(z_{\text{f}}) = 0.5\,M(z=0)$.  Though the
    formation redshift does not have a one-to-one correspondence with
    accretion type, Type~III halos tend to form early and Type~$0$
    halos tend to form late (cf. figure~\ref{fig:zf-contours}).}
  \label{fig:m-vs-z}
  \end{minipage}
\end{figure*}
As discussed in section~\ref{sec:method}, we do not solve for the
evolution of the dark matter; instead, we assume that the dark matter
evolves independently of the baryons and we parametrize it using fits
to halos from cosmological n-body simulations.  In particular, we
assume that the dark matter follows an \nfw distribution at all times
\citep{Navarro1997}, with a constant concentration parameter $c=5$, as
is appropriate for actively forming, massive halos \citep{Zhao2009}.
The only free parameter in this model for the dark matter is the mass
accretion history of the halo.

We use fits to halo mass accretion histories of the form
\begin{align}
  m(z) = [(1+z)^{b} \exp(-z)]^{\gamma} ,\label{eq:mdot-james}
\end{align}
derived by \citet{McBride2009} from the Millennium simulation
\citep{Springel2005}.  This fit is calibrated to the ``friends of
friends'' mass $M_{\text{FoF}}$ \citep{Davis1985}, which is similar to
our choice of $M_{200}$ for the virial mass \citep{White2001}.
Figure~\ref{fig:zf-contours} shows the parameter space for the
exponents $\gamma$ and $b$, along with the definitions of Type~I, II,
III, and IV accretion histories from
\citet{McBride2009},\footnote{\citet{McBride2009} do not distinguish
  between Type~II halos with negative and positive values of the
  exponent $b$.  Since this distinction is important in our
  application, we denote Type~II halos with $b<0$ as Type~$0$.}  and
figure~\ref{fig:m-vs-z} shows an example of each ``Type'' of accretion
history.  For reference, the red curves in
\mfigure~\ref{fig:zf-contours} show contours of the probability
density function for halo accretion parameters from the appendix of
\citet{McBride2009}.  We also show contours of the ``formation
redshift'' $z_{\text{f}}$ at which the cluster reaches half its
present mass; most of the halos formed at a redshift between~$0.5$
and~$1$.

Note that a Type~I accretion history is roughly exponential in
redshift, which is typical in $\Lambda$CDM \citep{Wechsler2002}.  At
fixed mass, Type~$0$ halos are younger, while Type~II and Type~III
halos are older, than Type~I halos.  A Type~IV accretion history
corresponds to mass loss at late times; this would imply a negative
ram pressure at the virial shock in our quasi-equilibrium model.
These cases should be studied with fully dynamical, cosmological
simulations.  Fortunately, at the high masses we wish to study, a
relatively small fraction of the total halos exhibit Type~IV accretion
histories.  Moreover, these are systems which have recently undergone
major mergers; they are likely to be morphologically disturbed and may
be excluded from cosmological samples.

In order to calculate the strength of the virial shock, we require an
estimate for the turnaround radius $r_{\text{ta}}$
(\S~\ref{sec:method}).  Unfortunately, fits to $r_{\text{ta}}$ from
n-body simulations do not seem to be available.  Therefore, we simply
use the virial theorem to estimate that the turnaround radius is twice
the virial radius.  The shock radius is then given in terms of the
virial radius by $r_{\text{sh}} = 2 \xi r_{\text{vir}}$.  We note that
this approximation may over-estimate the turn-around radius
\citep[cf.][]{Diemand2007}, causing us to predict temperature profiles
which are too shallow.  This is one of the primary sources of
uncertainty in our models.

The quasi-equilibrium model for the dark matter described in this
section greatly simplifies our method by eliminating the need to solve
for the dark matter dynamics (\eg by using an n-body simulation or by
solving the Jeans equations).  This model is ambiguous outside the
virial radius, however.  Since the virial shock typically lies
exterior to the virial radius, it is not a priori clear what value of
$M_{\text{sh}}$ to use in equation~\ref{eq:shock-entropy} for the
post-shock entropy.  We proceed by presuming that the gas and dark
matter first separate at the virial shock; thus, the gravitating mass
$M_{\text{sh}}$ in equation~\ref{eq:shock-entropy} corresponds to the
mass $M_{\text{vir}}$.  After the shock, the gas remains in
hydrostatic equilibrium at $r_{\text{sh}}$ while the dark matter
continues to collapse and virializes at the virial radius
$r_{\text{vir}} < r_{\text{sh}}$.  Therefore, when we solve for
hydrostatic equilibrium in the post-shocked gas, we assume that the
dark matter has relaxed and we extrapolate the \nfw profile between
the virial- and shock radii.  This inconsistency in our treatment of
the gravitating mass is an unavoidable consequence of applying a
quasi-equilibrium model for the dark matter outside the virial radius.

\subsubsection{Gas Equations}
\label{subsubsec:conduction-gas}
We solve for the state of the gas using a method very similar to that
described in section~\ref{sec:adiabatic}, but we now solve for the
shock radius $\xi(m)$ self-consistently.  We discretize the halo
formation into the accretion of a finite number of shells and, for
each shell $m_i$, we solve the full eigenvalue problem for its shock
strength $\xi(m_i)$.  Thus, we simultaneously build up solutions for
the shock radius and for the temperature profile as functions of time.
As discussed above, we assume that the gas and dark matter first
separate at the virial shock.  The baryonic accretion rate at the
shock radius $r_{\text{sh}}$ is thus also proportional to
equation~\ref{eq:mdot-james}.  This is qualitatively consistent with
the findings of \citet{Faucher2011}, who show that the baryonic
accretion closely tracks the dark matter accretion history in
simulations of high-mass halos.

The equations for mass conservation and hydrostatic equilibrium are:
\begin{subequations}
\begin{align}
  \frac{d x}{d y} &=
    \frac{h^2}{x^2}
    \left(\frac{K_1(y)}{P_1(y)}\right)^{3/5} ,\label{eq:lagrangian-x}\\
  \frac{d P_1}{d y} &=
  -\frac{m^{2/3} h^{8/3}}{x^3}
    \frac{1}{x}
    \frac{\log(1+c x)-c x/(1+c x)}{\log(1+c)-c/(1+c)} ,\label{eq:lagrangian-p}
\end{align}\label{eq:nfw-system}
\end{subequations}
with the boundary conditions:
\begin{subequations}
\begin{align}
  P_1(y = 1) &= \frac{4\sqrt{2}}{3} m^{2/3} h^{8/3} \eta
                \sqrt{\frac{1-\xi}{(2 \xi)^5}} \label{eq:nfw-bc-p}\\
  x(y = 1) &= 2 \xi \\
  x(y = 0) &= 0.
\end{align}\label{eq:nfw-bcs}
\end{subequations}
In the above, $\eta \equiv \ts{dyn} (\partial \ln M_{\text{vir}}
/ \partial t)$ measures the accretion rate of the halo, and $h \equiv
H/H_0 = [\Omega_{\text{m}} (1+z)^3 + \Omega_{\Lambda}]^{1/2}$ is the
Hubble parameter in units of $H_0$.  In equations~\ref{eq:nfw-system}
and~\ref{eq:nfw-bcs}, $\eta$, $h$, and $\xi$ are all evaluated at the
epoch of the most recently accreted shell.  It is straightforward to
show from equation~\ref{eq:mdot-james} that $\dot{m} = m \gamma(1+z-b)
H$.  Thus, the dimensionless accretion rate $\eta$ is given by $\eta =
\gamma(1+z-b)/10$.

The shock entropy $K_1^{\text{sh}}$ takes the form:
\begin{align}
  K_1^{\text{sh}}(y) &= \frac{1}{3}
  \left[
    \frac{y}{\eta[z(y)] \times h[z(y)]}
  \right]^{2/3}
  [1-\xi(y)]^{4/3} ,\label{eq:cond-entropy}
\end{align}
where $z(y)$ represents the redshift at which the baryonic shell $y$
accreted and $\xi(y)$ represents the shock radius of the cluster at
that redshift.

Inserting the conductivity appropriate for a fully-ionized hydrogen plasma
\citep{Spitzer1962} into equation~\ref{eq:conduction} yields:
\begin{align}
\begin{split}
  \frac{d \ln K_1}{d m} = 0.70 & f_{\text{Sp}} \frac{4\pi}{\fb}
  \frac{\mu^{7/2}}{h m \eta}
  \left(\frac{M_{\text{vir},\,0}}{10^{15} M_{\odot}}\right) \\
  & \times K_1^{5/2}
  \left[
    \frac{2}{x} \partiald{T_1}{x}
    + \frac{5}{2 T_1} \left(\partiald{T_1}{x}\right)^2
    + \frac{\partial^2 T_1}{\partial x^2}
  \right] .
\end{split} \label{eq:lagrangian-conduction}
\end{align}
(Recall that the coordinate $m$ tracks the formation of the cluster
and thus functions as a time coordinate).  We integrate this equation
between accretion events with an explicit, sub-cycled method.  The
conductive heat flux must vanish at the origin by spherical symmetry;
thus, we adopt $\partial T/\partial r = 0$ as the inner boundary
condition for equation~\ref{eq:lagrangian-conduction}.  The precise
boundary condition on the heat flux at the shock is uncertain because
it depends on the physics of collisionless shocks in the presence of
strong thermal conduction.  We proceed by assuming that electrons do
not diffuse across the shock into the upstream flow; in our model, the
shock thus serves as an insulating boundary.  This assumption is
convenient because thermal conduction does not modify the structure of
the shock.  We have also tried calculations in which we keep the heat
flux constant at the shock; the results with this alternative boundary
condition were very similar to those we present here.

Note that equations~\ref{eq:nfw-system}--\ref{eq:cond-entropy} are
independent of halo mass, while
equation~\ref{eq:lagrangian-conduction} is not.  Conduction thus
introduces non-self-similar behavior and may influence mass-observable
relations.  We quantify this departure from self-similarity in the
following section.

The post-shock entropy (eq.~\ref{eq:cond-entropy}) and the outer
boundary condition on the pressure (eq.~\ref{eq:nfw-bc-p}) depend on
the accretion history through the dimensionless accretion rate $\eta$
(cf.~\S~\ref{sec:adiabatic}).  Thus, the diversity in accretion
histories may generate scatter in the \icm properties at fixed halo
mass.  In order to estimate the statistics in \icm properties, we
generate an ensemble of accretion histories for each halo mass
$M_{\text{vir},\,0}$, with the exponents $\gamma$ and $b$ drawn from
the distribution in the appendix of \citet{McBride2009}.  This
ensemble yields information about the statistics of the cluster
population and the extent to which the variation in accretion
histories creates scatter in the temperature profile and
mass-observable relations.  We present these results in the next
section.

\subsection{Results}
\label{subsec:conduction-results}
\begin{figure*}
  \begin{minipage}[t]{3.33in}
  \centering
  \includegraphics[width=3.33in]{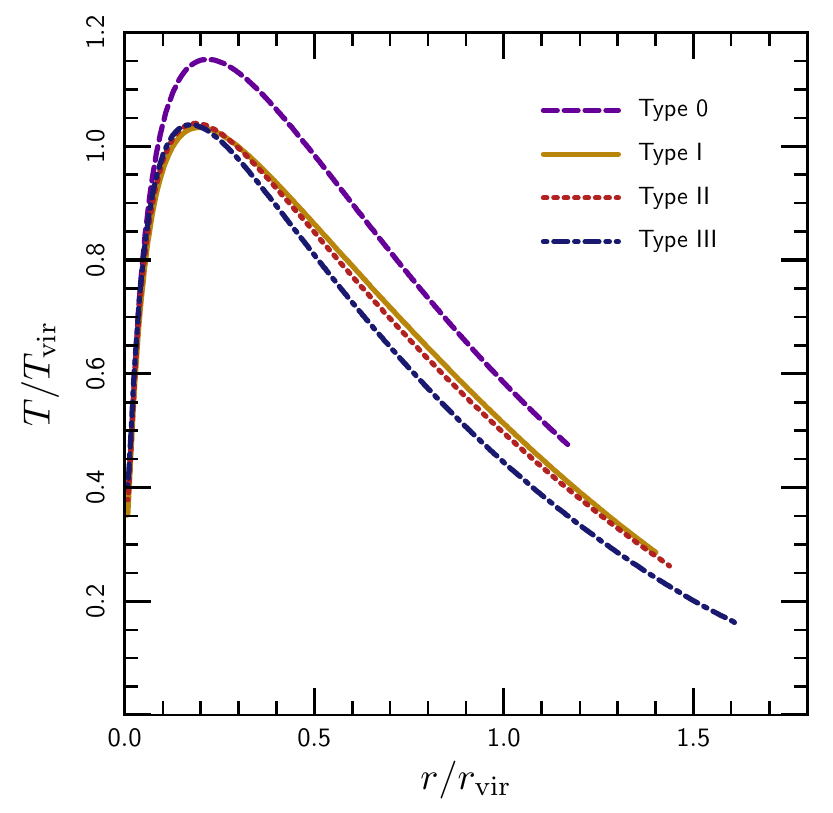}
  \caption{Representative temperature profiles at redshift $z=0$
    generated by each type of accretion history from
    \citet{McBride2009} (see figures~\ref{fig:zf-contours}
    and~\ref{fig:m-vs-z}).  The profiles are calculated using the
    Lagrangian method from section~\ref{sec:conduction}, but with no
    thermal conduction (the temperature profiles are thus independent
    of the halo mass).  Note the qualitative agreement with the simple
    models from section~\ref{sec:adiabatic}.  In addition, the
    location of the shock radius depends fairly sensitively on the
    accretion history of the halo.}%
  \label{fig:temp-by-type}
  \end{minipage}
  \hspace*{\fill}
  \begin{minipage}[t]{3.33in}
  \includegraphics[width=3.33in]{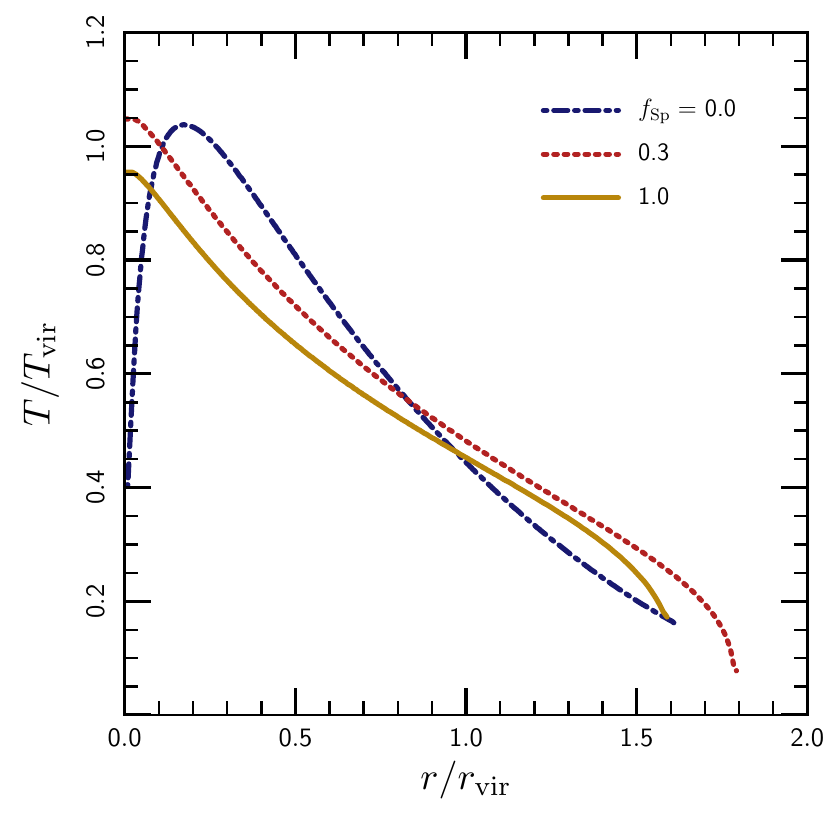}
  \caption{Influence of the effective conductivity on cluster
    temperature profiles.  These curves show temperature profiles for
    massive, 10\textsuperscript{15}$M_{\odot}$ halos with Type~III
    accretion histories at redshift $z=0$.  This choice of mass,
    accretion type, and redshift maximizes the effect of thermal
    conduction; nonetheless, the change to the temperature profile at
    large radii is modest.  As discussed in the text, the marked
    effect of thermal conduction on the temperatures at small radii is
    not a firm prediction, since our models neglect both radiative
    cooling and feedback heating, which are important in this region
    \citep{Voit2011}.}%
  \label{fig:temp-by-cond}
  \end{minipage}
\end{figure*}
Before studying the effect of thermal conduction, we present adiabatic
models in which we set the effective conductivity to zero.  These
profiles facilitate comparison with the simpler models in
section~\ref{sec:adiabatic} and also provide a basis for understanding
the models with conduction.

Figure~\ref{fig:temp-by-type} shows representative, adiabatic
solutions for the redshift-zero temperature profiles resulting from
several different accretion histories.  (Because we have neglected
thermal conduction, these solutions are independent of halo mass.)  In
all cases, the temperature approaches the virial temperature near the
scale radius of the halo, and falls by a factor of $\sim2$ by the
virial radius.  We note that the temperature profiles in
\mfigure~\ref{fig:temp-by-type} are not monotonic with density and
thus cannot be described with polytropic models.  Interestingly,
however, a polytrope with an index $\gamma \sim 1.2$ (as assumed by
\citetalias{Voit2003}) provides a good fit between the scale radius
and $\sim 0.5\,r_{\text{vir}}$.  These profiles are qualitatively very
similar to the \nfw models in
\mfigure~\ref{fig:isothermal-temp-profiles} with $\ts{dyn}/t_0 \sim
0.15$, as is reasonable for clusters
(\S~\ref{subsec:adiabatic-results}).  The location of the accretion
shock is also consistent with our interpretation in
section~\ref{sec:adiabatic}: Type~III halos, which experience slower
accretion at late times, have larger shock radii than the more rapidly
accreting Type~$0$ halos.  Thus, the intuition we developed in
section~\ref{sec:adiabatic} likely holds even for the more complex
models in this section.

Ignoring thermal conduction is not a well-motivated approximation,
however: as discussed in section~\ref{sec:introduction}, the timescale
for heat to diffuse through massive clusters
($r_{\text{vir}}^2/\chie\sim1$\,Gyr) is shorter than the typical age
of the \icm ($\sim5$\,Gyr).  Consequently, non-cosmological
simulations of isolated halos \citep[\eg][]{Parrish2008} show that the
\icm becomes almost completely isothermal after $\sim2$\,Gyr.  By
analogy, one might therefore expect conduction to significantly modify
the temperature profile shown in
\mfigure{s}~\ref{fig:isothermal-temp-profiles}
and~\ref{fig:temp-by-type}.

Figure~\ref{fig:temp-by-cond} compares the temperature profiles of
clusters with different effective conductivities.  In order to
maximize the influence of thermal conduction, we show
10\textsuperscript{15}$M_{\odot}$ clusters (which are hotter, and thus
more conductive than lower mass clusters), with Type~III accretion
histories (which formed comparatively early and thus provide more time
for conduction to operate).  As expected, thermal conduction smooths
out the temperature profile in the \icm; the effect, however, is
substantially weaker than has been found in non-cosmological
simulations (see above).  The profiles we obtain do \emph{not} become
isothermal, in qualitative agreement with x-ray observations of the
\icm \citep{George2009,Simionescu2011}, and also with cosmological
simulations of clusters which include thermal conduction
\citep{Dolag2004,Burns2010}.  The disagreement with simulations of
isolated halos suggests that cosmological accretion and the continued
formation of clusters is essential not only for understanding the
origin of the large-scale temperature gradient in the \icm
(\S~\ref{sec:adiabatic}), but also for how this gradient persists in
spite of thermal conduction.

Two important effects differentiate cosmological calculations of
thermal conduction from non-cosmological simulations of isolated
halos.  First, even though the present-day timescale for thermal
conduction is shorter than the mean age of massive clusters, the gas
near the accretion shock is always young and thus unaffected by
conduction.  The jump conditions at the virial shock therefore still
determine the outer temperature, even when we take thermal conduction
into account.  The second effect differentiating cosmological and
non-cosmological calculations is that the \icm at higher redshift had
a lower temperature and thus a much lower conductivity than it has at
redshift zero.  Even at small radii, the age of the \icm
($\sim5$\,Gyr) thus over-estimates the timescale over which thermal
diffusion operates.  These two effects, which are not present in
non-cosmological simulations, strongly limit the influence of thermal
conduction on the large-scale temperature profile of the \icm.

\begin{figure}
  \centering
  \includegraphics[width=3.33in]{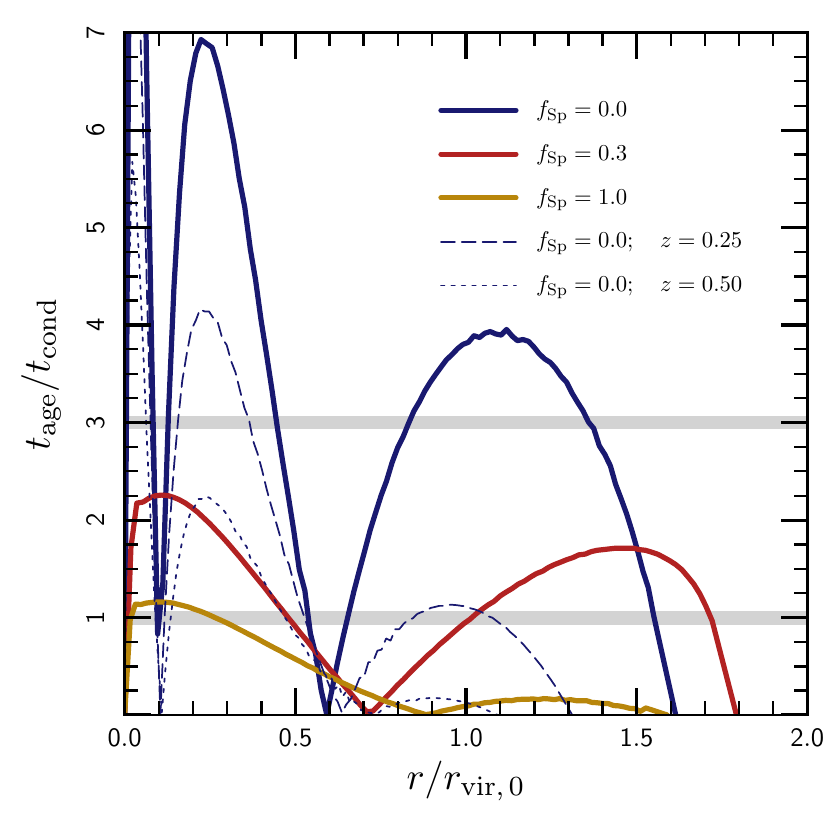}
  \caption{Ratio of the age of each shell in the \icm to its
    field-free conduction timescale (see text for definitions).  The
    models are the same as in \mfigure~\ref{fig:temp-by-cond}.  This
    ratio is defined using the field-free conductivity (\ie for
    $f_{\text{Sp}} = 1$).  Models with finite suppression factors
    should satisfy $\ts{age}/\ts{cond} \lesssim f_{\text{Sp}}^{-1}$;
    the thick gray lines show this limit for $f_{\text{Sp}} = 1$ and
    $1/3$.  Adiabatic models predict $\ts{cond} \lesssim (0.2-0.3)
    \ts{age}$ at intermediate radii; as a result, conduction alters
    the temperature profile and maintains $\ts{age}/\ts{cond} \lesssim
    f_{\text{Sp}}^{-1}$.}%
  \label{fig:rat-z}
\end{figure}
We illustrate these points in \mfigure~\ref{fig:rat-z}, where for each
model from \mfigure~\ref{fig:temp-by-cond} we plot the ratio of the
age of each shell in the \icm to its field-free conduction timescale.
We define the conduction timescale as $\ts{cond} \equiv |d\ln
K/dt|^{-1}$, calculated using equation~\ref{eq:conduction} with
$f_{\text{Sp}} = 1$.  We define the age of the shell \ts{age} as the
time over which conduction operates at its present-day efficiency; we
approximate this as the minimum of the time since the shell accreted
and $|d\ln m/dt|^{-1}$, the timescale over which the \icm conductivity
changes appreciably.  Conduction should limit the temperature profile
of the \icm to everywhere satisfy $\ts{cond}/\ts{age} \lesssim
f_{\text{Sp}}^{-1}$.  Thus, by looking at the adiabatic models in
\mfigure~\ref{fig:rat-z} (blue, with $f_{\text{Sp}} = 0$), we can
estimate how large of an effect conduction would have if we were to
include it.

\begin{figure*}
  \centering
  \includegraphics[width=7.0in]{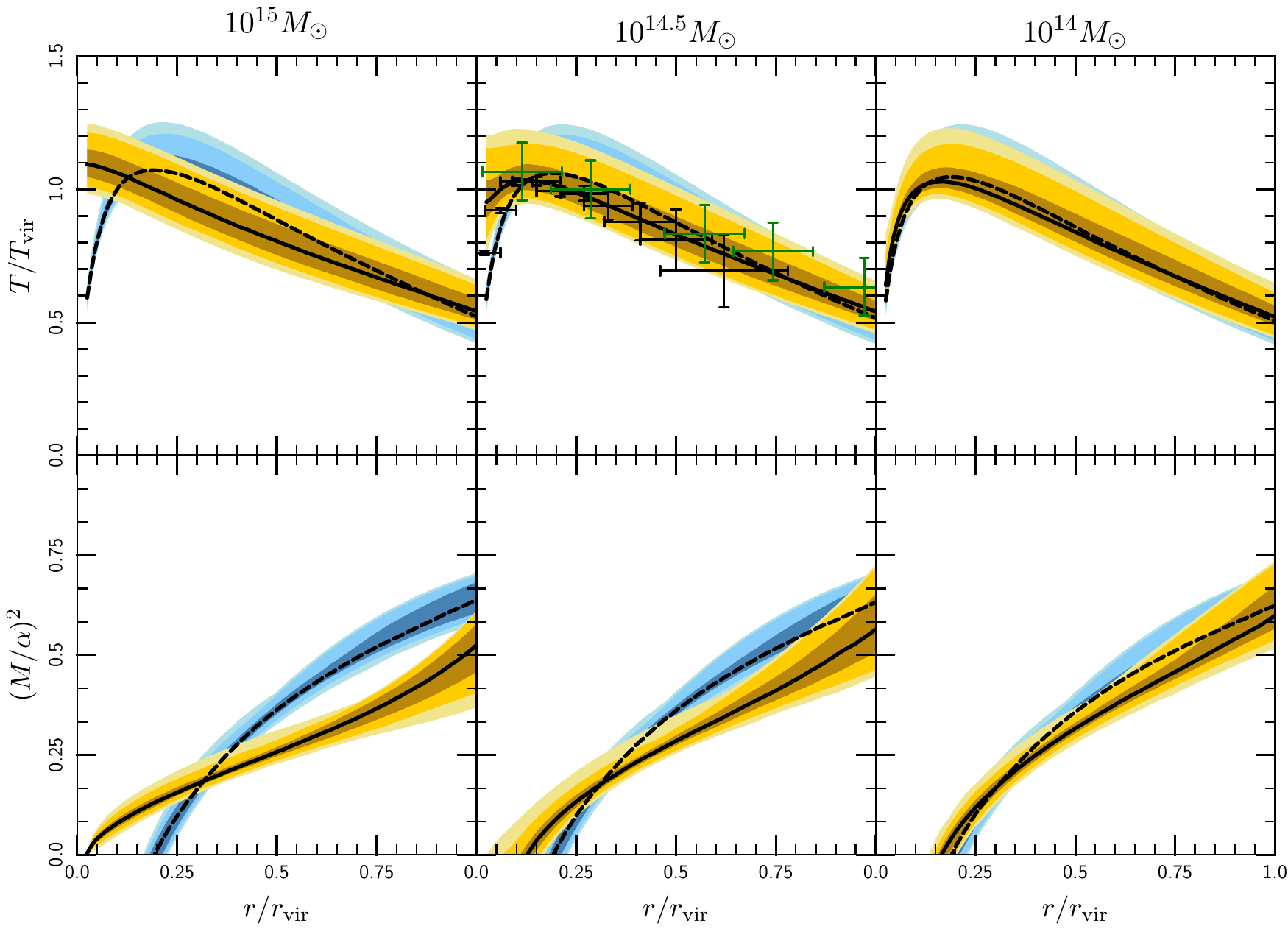}
  \caption{\emph{(Top row):} Temperature profiles with (yellow;
    $f_{\text{Sp}} = 0.3$) and without (blue) thermal conduction for
    different halo masses.  Black lines show the median profile from
    our ensemble of accretion histories based on
    \mfigure~\ref{fig:zf-contours} and colors show contours enclosing
    50\%, 90\%, and 99\% of the models.  The points with error bars in
    the top-right panel show temperature profiles derived from x-ray
    observations.  The black and green points are taken from
    \citet{Leccardi2008} and \citet{Simionescu2011}, respectively.
    \emph{(Bottom row):} Estimated fraction of non-thermal pressure
    support generated by the \mti.  See section~\ref{subsec:mlt} for
    details.}
  \label{fig:conduction-temp-mach}
\end{figure*}

Figure~\ref{fig:rat-z} shows that the ratio $\ts{age}/\ts{cond}$ falls
steeply near the virial shock, again reflecting that the recently
accreted gas is too young to be influenced strongly by conduction.  At
smaller radii, however, $\ts{age}/\ts{cond} \sim 5$ in the adiabatic
model, suggesting that thermal conduction plays a significant role in
determining the temperature profile in the \icm.  The thin blue curves
in \mfigure~\ref{fig:rat-z} show that this conclusion depends strongly
on redshift, however.  The flattening of the temperature profile shown
in \mfigure~\ref{fig:temp-by-cond} is thus a fairly recent phenomenon
and is far less pronounced at redshift $z \gtrsim 0.2$.  Recall also
that \mfigure{s}~\ref{fig:temp-by-cond} and~\ref{fig:rat-z} show
results for massive, 10\textsuperscript{15}$M_{\odot}$ clusters with
Type~III accretion histories.  Clusters with lower masses or different
accretion histories are likely to be even less strongly influenced by
conduction (see \mfigure~\ref{fig:conduction-temp-mach}, below).

The red and yellow curves in \mfigure~\ref{fig:rat-z} show that when
we include thermal conduction with a given suppression factor
$f_{\text{Sp}}$, the temperature gradient adjusts so as to keep the
ratio $\ts{cond}/\ts{age}$ below $f_{\text{Sp}}^{-1}$.  The ratio
$\ts{cond}/\ts{age}$ is proportional to the gradient term:
\begin{align*}
  \left[
    \frac{2}{x} \partiald{T_1}{x}
    + \frac{5}{2 T_1} \left(\partiald{T_1}{x}\right)^2
    + \frac{\partial^2 T_1}{\partial x^2}
  \right] .
\end{align*}
Thus, thermal conduction can either make the \icm isothermal (so that
the heat flux vanishes), or make $T \sim r^{-2/7}$ (so that the heat
flux is constant).  We find that our model clusters initially take the
second approach: the temperature profile adjusts so as to have a
nearly constant heat flux as a function of radius.  This is not
possible near the origin, however, where spherical symmetry requires
the heat flux to vanish.  Thus, over a longer timescale, the center of
the \icm cools and the \icm begins to become isothermal.  The yellow
curve in \mfigure~\ref{fig:temp-by-cond} shows the beginning of this
process, but there is not enough time before redshift $z=0$ for a
significant fraction of the \icm to become completely isothermal, even
in cases where conduction is most effective.

\begin{figure}
  \centering
  \includegraphics[width=3.33in]{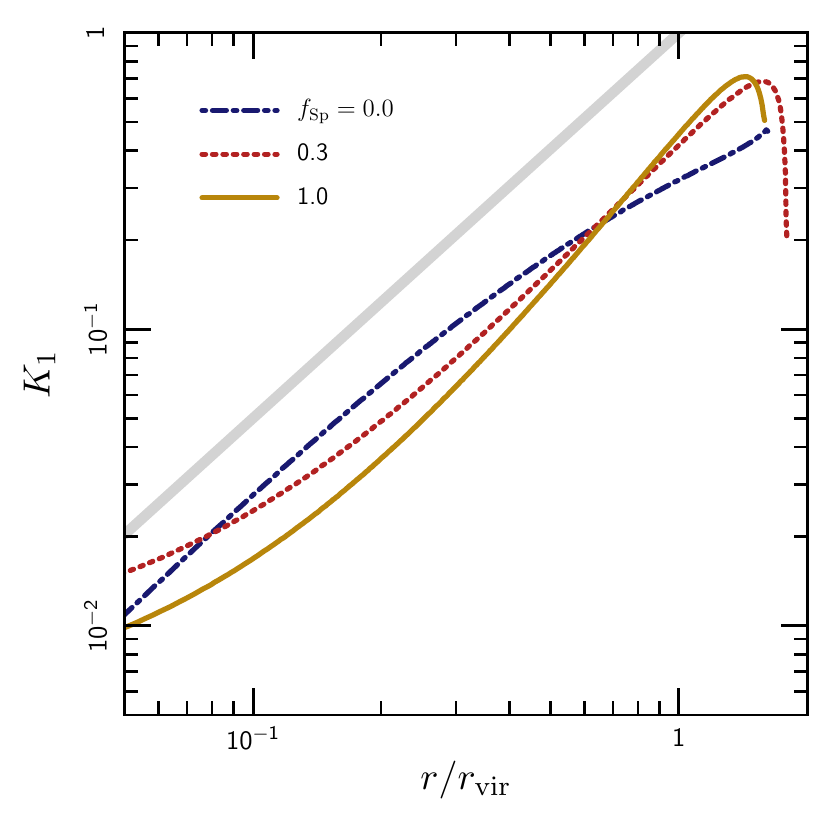}
  \caption{Entropy profiles in calculations with different values of
    the conductivity.  As in \mfigure~\ref{fig:temp-by-cond}, this
    figure shows 10\textsuperscript{15}\,$M_{\odot}$ halos with
    Type~III accretion histories; these properties maximize the effect
    of thermal conduction.  Also shown is the $K \sim r^{1.3}$
    power-law derived in \citetalias{Voit2003}.}%
  \label{fig:K-cond}
\end{figure}
Though conduction may have considerable effect on the temperature
profiles of massive clusters by redshift zero, its influence on their
entropy profiles is less pronounced.  We show this in
\mfigure~\ref{fig:K-cond}, which compares entropy profiles for
massive, 10\textsuperscript{15}$M_{\odot}$ clusters with and without
conduction.  Between the scale radius and the virial radius, both
models agree fairly well with a power-law fit derived from x-ray
observations \citep[\eg][]{Cavagnolo2009}.  Interestingly, thermal
conduction smooths out some of the non-power-law behavior introduced
by the accretion history, leading to an entropy profile more similar
to the adiabatic self-similar profile.

Figures~\ref{fig:temp-by-type}--\ref{fig:K-cond} illustrate the
effects of cosmological accretion and thermal conduction on the
temperature of the \icm in individual clusters.  As mentioned in
section~\ref{subsec:conduction-method}, in order to study the
statistics of temperature profiles as a function of virial mass, we
run an ensemble of models with accretion histories drawn from the
probability distribution in the appendix of \citet{McBride2009}.  The
upper panels of \mfigure~\ref{fig:conduction-temp-mach} show these
temperature profiles for three different halo masses with (yellow) and
without (blue) thermal conduction.  In the models with conduction, we
have assumed a conductive suppression factor $f_{\text{Sp}} = 0.3$.
The temperature profile roughly follows the pattern found in
section~\ref{sec:adiabatic}, but with fairly significant, $\sim$10\%,
scatter.  This scatter is caused purely by the variation in accretion
histories.\footnote{Note that we have neglected any correlation
  between halo concentration and accretion history, which may
  introduce additional variation in the temperature profiles
  \citep[\eg][]{Bullock2001}.  The analysis in \citet{Zhao2009}
  suggests that a constant value for the concentration $c\sim5$ is
  appropriate for massive clusters, however.} We have neglected
several processes, including mergers, heating by dynamical friction,
radiative cooling, and heating by \agn outflows, all of which may
increase the scatter above that shown here.

Figure~\ref{fig:conduction-temp-mach} also illustrates the
mass-dependence of conduction's role in the \icm.  Because the thermal
conductivity of a plasma depends sensitively on its temperature as
$\kappa \propto T^{5/2}$, higher-mass halos are more strongly
influenced by conduction.  More quantitatively, the conduction
timescale in the \icm falls with mass: $\ts{cond} \sim
r_{\text{vir}}^2/\chie \sim 1/M_{\text{vir}}$.
Figure~\ref{fig:conduction-temp-mach} shows that the normalization is
such that conduction has a minor influence on
10\textsuperscript{14}$M_{\odot}$ halos,\footnote{This conclusion
  applies to the large-scale temperature gradient in the \icm.  Of
  course, small-scale features may be strongly influenced by
  conduction, even in 10\textsuperscript{14}$M_{\odot}$ halos
  \citep[\cf][]{Dolag2004}.} but that it is significant for more
massive, 10\textsuperscript{15}$M_{\odot}$ clusters.

Finally, in the upper-right panel of
\mfigure~\ref{fig:conduction-temp-mach}, we have overlaid
observational data from \citet{Leccardi2008} and
\citet{Simionescu2011}.\footnote{The observation presented in
  \citet{George2009} provides another useful constraint on the
  properties of the \icm at large radii.  We do not include it in
  \mfigure~\ref{fig:conduction-temp-mach}, however, because it is not
  de-projected and thus not directly comparable to our results.}  The
data from \citet{Leccardi2008} (shown in black) represent an average
temperature profile derived from 48 clusters from the XMM-Newton
archive.  The points from \citet{Simionescu2011}, shown in green,
represent \textsc{suzaku} observations of the Perseus cluster out to
$r_{200}$.  Outside the scale radius, where our method is applicable,
our model agrees favorably with the data.  Though this agreement is
encouraging, we caution that both the normalization and slope of the
data points in this plot are sensitive to the assumed virial mass of
the halo, which is uncertain in x-ray observations.  Furthermore,
several uncertainties in our model, including the turnaround radius,
the effect of non-smooth accretion, and deviations from spherical
symmetry preclude a very quantitative comparison with the data.

\subsection{Mixing-Length Theory and the MTI}
\label{subsec:mlt}
As an example application of our models, we use mixing-length theory
to estimate the turbulent pressure support produced by convection in
clusters.  In dilute, magnetized plasmas such as the \icm in galaxy
clusters, convective stability depends on the temperature gradient of
the plasma (\citealt{Balbus2001}, later generalized by
\citealt{Quataert2008} and \citealt{Kunz2011}).  This convection,
known as the \mti, may produce strong turbulence in clusters
\citep{McCourt2011,Parrish2012} and thus may provide enough
non-thermal pressure support to bias hydrostatic mass estimates of
cluster halos.  \citet{Parrish2012} found that the convective
velocities produced by the \mti roughly obey mixing-length theory.
This motivates us to use our model temperature profiles to estimate
the turbulent pressure support produced by the \mti as a function of
cluster mass and redshift.  We note, however, that \citet{Kunz2012}
have shown that the strength of this turbulence depends on the
magnetic field strength in the \icm, an effect which we do not account
for in our simple estimates.

Assuming that the convective motions retain their coherence for a
fraction $\alpha$ of a pressure scale-height and that magnetic tension
does not suppress the convective motions, we expect the instability to
drive turbulent convection with Mach numbers of order:
\begin{align}
  M \sim \alpha \left(H \frac{d \ln T}{d r}\right)^{1/2} ,\label{eq:mti-mlt}
\end{align}
where $H \equiv (d \ln P / d r)^{-1}$ is the pressure scale-height.
The bottom panels of \mfigure~\ref{fig:conduction-temp-mach} show this
estimate for the turbulent pressure support (proportional to $M^2$).
Assuming (as suggested by the simulations in \citealt{Parrish2012})
that the mixing length parameter $\alpha\sim0.5$, this \mfigure
suggests that turbulence driven by the \mti contributes $\sim5-10\%$
of the pressure support outside of $r_{500}$.  Cosmological
simulations of cluster formation
\citep[\eg][]{Rasia2006,Nagai2007,Vazza2011} find significant
turbulent pressure support due to subsonic, bulk flows driven by
mergers near the virial radius.  We note that any turbulence produced
by the \mti would add to, but would likely be sub-dominant to, that
produced by infalling subhalos.

Since the \mti is not suppressed by other sources of turbulence
\citep{McCourt2011,Parrish2012}, this pressure support adds to that
already present due to turbulence driven by infalling substructure or
by galaxy wakes.  This analysis suggests that the \mti plays an
important role in the dynamics of the \icm, especially at radii
$\gtrsim 0.5 r_{\text{vir}}$.  Thus, the \mti may provide an
interesting correction for hydrostatic mass estimates of cluster
halos.  Unfortunately, both the large scatter in the strength of the
\mti and its strong radial dependence \citep[cf.][]{Parrish2012} seem
to preclude a simple fitting function for the fraction of turbulent
pressure support as a function of halo mass.

By applying mixing-length theory in equation~\ref{eq:mti-mlt}, we have
implicitly assumed that the \mti grows rapidly enough to establish
convection by redshift zero.  We find that $\ts{\mti} \lesssim 0.5
\ts{age}$ within $r_{\text{vir}}$; thus, the Mach numbers in
\mfigure~\ref{fig:conduction-temp-mach} (and also in
\citealt{Parrish2012}) are likely to be reasonable estimates.

Another assumption implicit in our use of mixing-length theory is that
conduction is rapid enough to sustain the \mti.  Thus, the results in
the lower panels of \mfigure~\ref{fig:conduction-temp-mach} are only
valid when the conduction timescale across an unstable mode $\sim
(\alpha H)^2/\chi$ is less than the growth time of the \mti.  Massive
clusters ($M_{\text{vir}} \gtrsim 10^{14.5}~M_{\odot}$), likely
satisfy this ordering of timescales, but lower masses halos
($M_{\text{vir}} \lesssim 10^{14}~M_{\odot}$) may not.  Thus, the
mass-dependence of the \mti is not likely to be monotonic.  In the
most massive halos ($M_{\text{vir}} \gtrsim 10^{15}~M_{\odot}$),
thermal conduction is more efficient and weakens the temperature
gradient at intermediate radii (\mfig~\ref{fig:conduction-temp-mach}).
In lower mass halos ($M_{\text{vir}} \lesssim 10^{14}~M_{\odot}$), on
the other hand, conduction may not be fast enough to drive the \mti to
its full potential.  This analysis suggests that $10^{14.5}~M_{\odot}$
halos experience the most vigorous convection driven by the \mti.

\section{Discussion}
\label{sec:discussion}
This paper provides a simplified, spherically symmetric model for the
temperature profiles of the hot plasma in galaxy groups and clusters.
Our model is similar in spirit to earlier studies of the entropy
profiles in clusters \citep[\eg][]{Tozzi2001,Voit2003}, but builds on
these earlier studies by focusing on temperature and by including the
effects of thermal conduction.  Our results agree reasonably well with
the profiles derived from x-ray observations
\citep[cf.][]{Leccardi2008} and from numerical simulations
\citep[\eg][]{Dolag2004}.

We have shown that the large-scale temperature gradient in the \icm is
primarily determined by the accretion history of its halo: while the
gas near the center of a cluster reaches the virial temperature of the
halo, the temperature at the virial shock is determined by the ram
pressure of the accreting gas.  This difference sets the overall shape
of the temperature profile.  The timescale for thermal conduction
($\ts{cond} \sim r_{\text{vir}}^2/\chie$) is somewhat shorter than the
age of the \icm.  However, the influence of thermal conduction on the
global temperature profile in clusters is mitigated for two reasons.
The gas near the center of the cluster is less strongly effected by
conduction because it was cooler in the relatively recent past.  The
gas near the virial shock, on the other hand, has only recently
accreted and is younger than the conduction timescale.  Thus,
conduction has a diminished effect both near the center of the cluster
and near the outskirts.  As a result, it does not dramatically change
the mean temperature profile (\mfig~\ref{fig:conduction-temp-mach}).
Of course, conduction can have a dramatic effect on small-scale
inhomogeneities in the \icm \citep[cf.][]{Dolag2004}; such
inhomogeneities cannot be studied in our one-dimensional model.

Our results demonstrate the close relationship between the temperature
gradient in clusters and the cosmological evolution of the host halo.
This implies that numerical studies of isolated cluster models
\citep[\eg][]{Parrish2008} cannot correctly predict the evolution of
the large-scale temperature profile, though they are very useful for
studying other aspects of the \icm, such as the interplay among
cooling, feedback, and plasma instabilities within the scale radius of
the halo.
                         
One of our motivations for studying the effects of conduction and halo
accretion history on temperature gradients in clusters is that the
free energy in the non-zero temperature gradient drives an efficient
convective instability, the \mti.  The results in
section~\ref{subsec:mlt} (\eg \mfig~\ref{fig:conduction-temp-mach})
show that the turbulent pressure support generated by the \mti may be
of order $\sim5$ percent of the thermal pressure, and that it scales
non-monotonically with halo mass.  The magnitude of the turbulent
pressure support is sensitive to the accretion history and does not
seem amenable to a simple fitting formula.

Halo accretion histories have been studied extensively with numerical
simulations.  We use the fits to the Millennium simulation from
\citet{McBride2009} to estimate the scatter in temperature profiles as
a function of halo mass; at redshift $z=0$, this scatter is of order
10\%.  This scatter likely contributes to the dispersion in cluster
mass-observable relations relevant to x-ray and \sz observations.
Perhaps more interesting are the effects of thermal conduction and
convection, which introduce systematic changes to the temperature
profile with mass.  In particular, conduction smooths out the
temperature profile (and decreases the peak temperature in the halo)
by an amount that increases monotonically with halo mass
(\mfig~\ref{fig:conduction-temp-mach}); convection, on the other hand,
produces turbulent pressure support that is non-monotonic in halo
mass, peaking around 10\textsuperscript{14.5}$M_{\odot}$ halos
(\S~\ref{subsec:mlt}).

Figures~\ref{fig:temp-by-type}, \ref{fig:temp-by-cond},
and~\ref{fig:conduction-temp-mach} demonstrate that the effect of
thermal conduction on a cluster's temperature profile is at least as
large as the differences produced by normal variation in accretion
histories.  Any variation in the factor $f_{\text{Sp}}$ (which
parameterizes the suppression of the effective radial thermal
conductivity relative to the field-free value), if it exists, would
create additional scatter in the temperature profiles at fixed mass.
Possible effects influencing $f_{\text{Sp}}$ include magnetic draping
around infalling or orbiting substructure
\citep{Dursi2008,Pfrommer2010} and the strength of the magnetic field.
These processes may also contribute to the scatter in cluster
mass-observable relations.

The models presented in this paper provide a simple explanation for
the physics that sets the temperature profiles in galaxy groups and
clusters at large radii.  Our results are consistent with current
observational constraints on cluster temperature profiles at large
radii (\mfig~\ref{fig:conduction-temp-mach}).  They also highlight
several processes which may bias hydrostatic mass estimates of
clusters, including modifications to the scaling relation $T(M)$ by
conduction and by turbulence driven by the \mti.  The approximations
in this paper, especially the assumptions of smooth accretion and
spherical symmetry, preclude precise estimates of the non-thermal
pressure support produced by the \mti.  These limitations can be
addressed using cosmological simulations.

\acknowledgments
\noindent E.~Q., I.~P., and M.~M. were partially supported by NASA ATP
grant NNX$10$AC$95$G, Chandra theory grant TM$2$-$13004$X, and a
Simons Investigator award from the Simons Foundation to EQ.  We are
grateful to Chung-Pei Ma, James McBride, and Claude-Andr\'e
Faucher-Gigu\`ere for helpful discussions about the structure of
dark-matter halos and their mass accretion histories.  We also thank
Mark Voit for pointing out applications of the $T(M)$ relationship,
and we thank the anonymous referee for suggesting several improvements
to the original manuscript.  We made our figures using the open-source
program \textsc{Tioga}.  This research has made extensive use of
NASA's Astrophysics Data System. \\

\bibliography{outer-parts}

\end{document}